\documentclass[aps,prb,twocolumn,nofootinbib,showpacs,noshowkeyws]{revtex4}

\usepackage{amsmath,bm,amssymb,amscd}
\usepackage[USenglish]{babel}

\begin{document}

\title{$T>0$ Ensemble State Density Functional Theory Revisited}

\author{Helmut Eschrig}

\email{h.eschrig@ifw-dresden.de}
\homepage{http://www.ifw-dresden.de/~/helmut}

\affiliation{IFW Dresden, PO Box 270116, D-0111171 Dresden, Germany}

\begin{abstract}
  A logical foundation of equilibrium state density functional theory in
  a Kohn-Sham type formulation is presented on the basis of Mermin's
  treatment of the grand canonical state. it is simpler and more
  satisfactory compared to the usual derivation of ground state theory,
  and free of remaining open points of the latter. It may in particular
  be relevant with respect to cases of spontaneous symmetry breaking
  like non-collinear magnetism and orbital order.
\end{abstract}


\maketitle

\section{Introduction}

Modern ground state density functional theory (DFT) for an inhomogeneous
system of identical particles, having early roots in the work of Thomas
and Fermi, was pioneered by the seminal papers by Hohenberg, Kohn and
Sham.\cite{hk64,ks65} It was later generalized by the constrained search
concept of Levy\cite{lev82} and finally put on a mathematically rigorous
basis of functional Legendre transforms by Lieb.\cite{lie83} Meanwhile,
DFT for the quasi-particle self-energy labeled by acronyms like GW or
LDA+DMFT, and time-dependent DFT for dynamical processes, related to
Keldysh Green's functions, where developed, and all these theories layed
the ground for enormously successful model approaches (by use of model
functionals) to simulation of molecules and solids of any complexity.

A generalization of the ground state DFT to thermodynamic equilibrium
states by Mermin\cite{mer65} which appeared shortly after the work of
Hohenberg, Kohn and Sham, has been mentioned from time to
time,\cite{sto88,ww92} but otherwise has been largely ignored so far.

As it sometimes happens,\cite{kl60} a formalism for temperature $T=0$
need not be equivalent to that for $T \downarrow 0$ which latter case is
always the relevant case in physics. Although ground state DFT is
largely settled now, some uneasy feeling remains in connection with the
density functional failing to be differentiable in some cases (where
$n_{ss'} \mapsto v_{ss'}$ is not unique), notably in spin
DFT.\cite{ep01,koh04} In Ref.~\onlinecite{ep01} it is argued that these
problems may reduce to the ordinary well understood gap problem, now for
the spin subsystems separately, if one restricts consideration to
homogeneous external magnetic fields only. This might seem a reasonable
restriction since applied fields in laboratory can hardly vary over
microscopic distances. However, in the very topical cases of spontaneous
symmetry breaking with respect to the interplay of non-collinear
magnetism with orbital order, in a statistical treatment one has to
resort to the trick of Bogolubov's quasi-means by applying a suitable
infinitesimal symmetry-breaking external field, otherwise statistical
ensembles would not reproduce the broken symmetry. In the just mentioned
cases this implies a microscopically inhomogeneous field, and one would
like to rely on a situation where everything is fine at least in an
infinitesimal vicinity of such a symmetry-breaking field.  The good news
is that this is indeed the case and the needed functional derivatives
always exist for $T>0$. This is shown with the help of Mermin's approach
in the sequel. DFT is a rigorous theory for volume $V<\infty$ and for
temperature $T>0$. For $V = +\infty$ a ground state wave function may
not exist and for $T = 0$ the functional derivative of the density
functional may not exist. The theory then may be applied for $V
\rightarrow \infty$ (as is routinely done with refinement of the grid in
$\bm k$-space; a discrete regular $\bm k$ grid means periodic boundary
conditions with a finite periodicity volume) and for $T \downarrow 0$.
(Contrary to the case of adiabatic molecular dynamics\cite{ww92} the
temperature of a finite system in an equilibrium state has a well
defined meaning in the average over states the system may be in after in
had been for a long time in contact with a large thermal bath.)

\section{Subtleties of Hohenberg-Kohn-Sham Theory}

Originally\cite{hk64} DFT was built for systems in an (arbitrarily
large) box of finite volume which conveniently can be replaced by
periodic boundary conditions meaning to treat the position space of the
particles as a three-torus $\mathbb T^3$ of finite volume $|\mathbb
T^3|$, for instance $x \equiv x+L,\; y \equiv y+L,\; z \equiv z+L,\;
|\mathbb T^3| = L^3$. 

Let the Hamiltonian be
\begin{equation}
  \label{eq:II01}
  \hat H = \hat T + \hat W + \hat V,
\end{equation}
\begin{equation}
  \label{eq:II02}
  \hat T = \frac{\hbar^2}{2m} 
  \int \nabla\hat\psi^\dagger(x)\nabla\hat\psi(x)\, dx,
\end{equation}
\begin{equation}
  \label{eq:II03}
  \hat W = \int \hat\psi^\dagger(x)\hat\psi^\dagger(x')
  w(\bm r,\bm r')\hat\psi(x')\hat\psi(x)\,dx'dx,
\end{equation}
\begin{equation}
  \label{eq:II04}
  \hat V = \int 
  \hat\psi^\dagger(\bm r,s)v(\bm r)\hat\psi(\bm r,s)\, dx,
\end{equation}
where $\hat\psi(x),\; x = (\bm r,s)$ with particle position $\bm r$
and spin variable $s$, is the field operator of the particle field and
$\int dx = \sum_s \int d^3r$.

Then, given any particle number $N$, a normalized ground state (GS)
wavefunction (WF) $\Psi(x_1,\ldots,x_N)$ and a GS density $n(\bm r)$
exist for any reasonable external potential $v(\bm r)$ and for any
non-negative pair interaction $w(\bm r_i,\bm r_j)$. As is now
standard,\cite{lie83} one allows for all potentials with the only
condition that $\int_{\mathbb T^3} |v|^{3/2}\, d^3r < \infty$, that is,
$v \in \bm L^{3/2}(\mathbb T^3)$. Potentials of arrays of finitely many
point charges in the $\mathbb T^3$ belong to this
space\cite{lie83,esc03}, and the Hamiltonian $\hat H^0 = \hat T + \hat
V$ of interaction-free fermions is bounded below for any such potential.
Then, this also holds true for Hamiltonians (\ref{eq:II01}), if $w(\bm
r,\bm r') \geq 0$. Since the space $\mathbb T^3$ has finite volume, all
considered Hamiltonians have discrete spectra with at most finite
degrees of level degeneracy.  (Lieb allowed the position space to be the
real vector space $\mathbb R^3$ of infinite volume which caused many
problems with the continuous part of the spectrum of Hamiltonians.  He
then had to restrict $n \in \bm L^3(\mathbb R^3) \cap \bm L^1(\mathbb
R^3)$ since the density must integrate to a finite particle number $N$
over the infinite space $\mathbb R^3$. This led him allow for potentials
$v \in \bm L^{3/2}(\mathbb R^3) + \bm L^\infty(\mathbb R^3)$.  In the
three-torus every function $n \in \bm L^3(\mathbb T^3)$ may be
normalized to integrate to a given $N$.)

The lemma by Hohenberg and Kohn\cite{hk64} states the unique mapping $n
\mapsto v$ from ground state densities (degenerate ground states
allowed\cite{lie83}) to external potentials, on which basis the
Hohenberg-Kohn density functional
\begin{equation}
  \label{eq:II05}
  F_{\text{HK}}[n] = E[v[n],N] - \langle n,v[n] \rangle
\end{equation}
is defined \textit{for any ground state density} $n \in {\cal A}_N$:
\begin{equation}
  \label{eq:II06}
  {\cal A}_N = 
  \{n \text{ comming from an $N$-particle GS-WF}\}.
\end{equation}
In (\ref{eq:II05}) $v[n]$ means the potential causing a WF-GS density $n$,
and henceforth we use the notation of linear functionals
\begin{equation}
  \label{eq:II07}
  \langle n,v \rangle = \int_{\mathbb T^3} n(\bm r)v(\bm r)\, d^3r.
\end{equation}
As is easily seen, $\langle n,v \rangle$ cancels an equal term in the GS
energy $E[v[n],N]$, so that $F_{\text{HK}}$ does not any more depend on
$v[n]$. The functional $F = F_{\text{HK}}$ might be used in the
variational principle by Hohenberg and Kohn
\begin{equation}
  \label{eq:II08}
  E[v,N] = \min_n \{F[n] + \langle v,n \rangle \,|\, 
  \langle 1,n \rangle = N \}
\end{equation}
where $\{A|B\}$ means a set of elements $A$ with property $B$. The
crucial point for the possibility to solve this problem with the help of
Euler's equation is the know\-ledge of the variational domain for $n$
and the existence of the functional derivative of $F$. Would the
derivative of $F$ at the minimizing density $n_{\text{min}}$ exist, it
would be $\delta F[n_{\text{min}}]/\delta n = -v + \mu$, where $\mu$ is
the Lagrange multiplier for the constraint in (\ref{eq:II08}). For
$F_{\text{HK}}$, unfortunately neither the domain of definition ${\cal
  A}_N$ is explicitly known nor is anything known about the existence of
the functional derivative. We know that ${\cal A}_N \subset \bm
L^3(\mathbb T^3)$, but Lieb has shown\cite{lie83} that ${\cal A}_N$ is
not convex: There are densities $n = \sum_i c_i n_i,\; c_i \geq 0,\;
\sum_i c_i = 1,$ which are not in ${\cal A}_N$ while the $n_i$ all are
in ${\cal A}_N$. This is, why nowadays more general definitions of
$F[n]$ are used.

As the theory can be build for any reasonable pair interaction $w$, the
interaction-free case $w = 0$ is of some help. Further on this case will
be denoted by a superscript 0, but the corresponding density functionals
$F^0$ will as usually be denoted by $T$ since they obviously reduce to
the kinetic energy of an interaction-free system with GS density $n$. In
this case, an alternative to (\ref{eq:II05}) is the density matrix (DM)
functional 
\begin{equation}
  \label{eq:II09}
  \begin{split}
  \lefteqn{T_{\text{DM}}[n] =}\\
  &= \min_{\substack{\{p_k,\varphi_k\}\\
     0 \leq p_k \leq 1\\
     (\varphi_k|\varphi_{k'}) = \delta_{kk'}}} 
  \biggl\{\sum_k p_k (\varphi_k|\hat t|\varphi_k) \biggm|
  \sum_k p_k |\varphi_k|^2 = n \biggr\}
  \end{split}
\end{equation}
with $\hat t = -\hbar^2\Delta/2m$. For any $N$-particle DM
state (not only GS),
\begin{equation}
  \label{eq:II10}
  \begin{split}
  &{\cal T}[\{p_k,\varphi_k\}] = 
  \sum_k p_k (\varphi_k|\hat t|\varphi_k),\\
  &0 \leq p_k \leq 1,\; \sum_k p_k = N,\;
  (\varphi_k|\varphi_{k'}) = \delta_{kk'},
  \end{split}
\end{equation}
is the general expression of the kinetic energy (with the set of
orthonormal orbitals $\varphi_k$ depending on the state). Given a
potential $v$, the orbitals with $(\hat t + v)\varphi_k =
\varphi_k\varepsilon_k$ and occupation numbers $p_k = 1$ for
$\varepsilon_k < \varepsilon_N$, $p_k = 0$ for $\varepsilon_k >
\varepsilon_N$ minimize (\ref{eq:II09}) for the corresponding GS density
$n \in {\cal A}^0_{\text{DM},N}$. The GS is unique if the highest
occupied level $\varepsilon_N$ is not degenerate. The GSs and their
energies are obtained from the Kohn-Sham (KS) variational principle
\begin{equation}
  \label{eq:II11}
  E^0[v,N] = 
  \min_{\substack{\{p_k,\varphi_k\}\\
  0 \leq p_k \leq 1, \sum p_k = N\\
  (\varphi_k|\varphi_{k'}) = \delta_{kk'}}} 
  \biggl\{{\cal T}[\{p_k,\varphi_k\}] +
  \sum_{k=1}^N (\varphi_k|v|\varphi_k) \biggr\}
\end{equation}
while (\ref{eq:II09}) is defined for the density $n$ of any $N$-particle
state, that is,\cite{lie83} on the domain
\begin{equation}
  \label{eq:II12}
  {\cal J}_N = \bigl\{n \,\bigm|\, 
  n(\bm r) \geq 0,\; \nabla (n^{1/2}) \in \bm L^2(\mathbb T^3),\; 
  \langle 1,n \rangle = N \bigr\}
\end{equation}
which is a convex subset of $\bm L^3(\mathbb T^3)$. (It has been
shown\cite{lie83} that a minimum (\ref{eq:II09}) exists for every $n \in
{\cal J}_N$.)

Let for the sake of simplicity $v$ have a non-degenerate GS (which is a
single determinant of orbitals in this case) with density $n$. It is
easily seen that the most general variation permitted by (\ref{eq:II09})
is a linear combination of $\{\delta\varphi_k =
\lambda_k\varphi_{l_k}\},\; \varepsilon_k \leq \varepsilon_N,\;
\varepsilon_{l_k} > \varepsilon_N,\; \lambda_k \rightarrow 0$. To lowest
order in the $\lambda_k$ it yields $\delta T_{\text{DM}} =
2\text{Re}\,\sum_k \lambda_k (\varphi_{l_k}|\hat t|\varphi_k) =
2\text{Re}\,\sum_k \lambda_k (\varphi_{l_k}|v|\varphi_k) = \langle
v,\delta n \rangle$. These variations reach every single determinant
state in a neighborhood of the considered GS (with respect to the $\bm
H^1(\mathbb T^{3N})$-norm $\|\Psi\|^2 = \int (|\Psi|^2 +
|\nabla\Psi|^2)\,d^{3N}r$).  Since\cite{lie83} single determinant states
map \textit{onto} ${\cal J}_N$, the corresponding $\delta n$ is a
general variation in a neighborhood of $n$ in ${\cal J}_N$ (relative to
the $\bm L^3(\mathbb T^3)$-norm), and hence the functional derivative of
the convex functional $T_{\text{DM}}$ exists at $n$ as a (Frechet)
derivative in ${\cal J}_N$ and equals $v$. The argument can be
generalized to the case of a degenerate GS, that is, for all $n \in
{\cal A}^0_{\text{DM},N}$.

Now, let $n \in {\cal J}_N\setminus{\cal A}^0_{\text{DM},N}$. Such
densities exist, for instance densities having nodes cannot be in ${\cal
  A}^0_{\text{DM},N}$. Assume that the derivative of $T_{\text{DM}}$
exists at that $n$. This means that there is some $u$ with $\delta
T_{\text{DM}} = \langle u,\delta n \rangle$ for all permitted $\delta
n$. Since $T_{\text{DM}}$ was shown\cite{lie83} to be convex, the
assumption implies that $n$ minimizes $T_{\text{DM}}[n'] - \langle u,n'
\rangle$ and hence is a GS density to the potential $-u$ in
contradiction to the presupposition. $T_{\text{DM}}$ \textit{has nowhere
  outside of} ${\cal A}^0_{\textrm{DM},N}$ \textit{a functional
  derivative.}

KS theory in the interacting case $w > 0$ now uses the splitting
\begin{equation}
  \label{eq:II13}
  F[n] = T_{\text{DM}}[n] + E_{\text{H}}[n] + E_{\text{XC}}[n],
\end{equation}
which defines the density functional $E_{\text{XC}}[n]$ through the
preceding ones. While this definition is correct, nothing can be said on
the existence of the functional derivative of $E_{\text{XC}}[n]$ for GS
densities $n \in {\cal A}_{\text{DM},N}$ since we do not know the sets
${\cal A}^0_{\text{DM},N}$ and ${\cal A}_{\text{DM},N}$ and cannot
assume ${\cal A}_{\text{DM},N} \subset {\cal A}^0_{\text{DM},N}$. Even
though $F$ has a functional derivative for $n \in {\cal
  A}_{\text{DM},N}$ (see below), $E_{\text{XC}}[n]$ can only have one
there if $T_{\text{DM}}$ has one.

Like in (\ref{eq:II11}), with $n = \sum_k p_k |\varphi_k|^2$ a KS
variational principle is set out with the KS equation as the
corresponding Euler equation. Now one may assume
\begin{equation}
  \label{eq:II14}
  \begin{split}
  \lefteqn{T_{\text{DM}}[n] + E_{\text{XC}}[n] =}\\
  &= \min_{\substack{\{p_k,\varphi_k\}\\
        0 \leq p_k \leq 1\\(\varphi_k|\varphi_{k'}) = \delta_{kk'}}}
  \biggl\{\sum_{k=1}^N p_k (\varphi_k|\hat t|\varphi_k) + 
  {\cal E}_{\text{XC}}[\{p_k,\varphi_k\}] \biggm|\\[-3ex]
  &\qquad\qquad\qquad\qquad\qquad\qquad\;\;
  \biggm| \sum_{k=1}^N p_k |\varphi_k|^2 = n \biggr\}
  \end{split}
\end{equation}
leading to a KS equation with a (non-linear) exchange and correlation
potential operator
\begin{equation}
  \label{eq:II15}
  \frac{1}{p_k}\,
  \frac{\delta{\cal E}_{\text{XC}}}{\delta\varphi^*_k(\bm r)} =
  \hat v_{\text{XC}}\varphi_k(\bm r), \quad
  (\hat v_{\text{XC}}\varphi_k)^* = \hat v_{\text{XC}}\varphi^*_k\,,
\end{equation}
if one assumes that ${\cal E}_{\text{XC}}$ depends on $p_k$ and on a
Hermitian form of the $\varphi_k$ only. Like in (\ref{eq:II09}), the
right hand side of (\ref{eq:II14}) might exist (not proven so far; only
${\cal E}_{\text{X}}[\{p_k,\varphi_k\}]$ is a simple explicitly known
expression). By inserting (\ref{eq:II14}) together with $n = \sum_k
p_k|\varphi_k|^2$ into (\ref{eq:II13}) it is seen that ${\cal
  E}_{\text{XC}}$ (if it exists) has derivatives with respect to $p_k$
and $\varphi_k$ since all the other terms in the equation have them
(even where the left hand side of (\ref{eq:II14}) has no derivative with
respect to $n$). Hence, $\hat v_{\text{XC}}$ will in general not be a
local potential function, it may in particular be orbital dependent
($\hat v_{\text{X}}$ is non-local, its orbital dependence is canceled by
inclusion of the orbital dependent self-interaction in both
$v_{\text{H}}$ and $v_{\text{X}}$; the model XC potentials with partial
self-interaction correction or in LDA+$U$ models are non-local and
orbital dependent). Would $v_{\text{XC}}$ exist as a local potential,
then the KS equation would always yield a solution $n \in {\cal
  A}^0_{\text{DM},N}$ and hence there would be ${\cal A}_{\text{DM},N}
\subset {\cal A}^0_{\text{DM},N}$ which can by no means be taken for
granted.

The only density functional $F$ for which the issue of the existence of
the functional derivative can be addressed in general is the Legendre
transform\cite{lie83} 
\begin{equation}
  \label{eq:II16}
  F[n] = \sup_{v \in X^*} \bigl\{E[v,N] - \langle n,v \rangle\bigr\}
\end{equation}
for both cases, $w = 0$ and $w \neq 0$.  It is convex and defined on the
whole functional space $X$ (it may take on the value $+\infty$ in part
of $X$), and if, given $n$, there exists a unique maximizing $v$, then
this is the functional derivative of $F[n]$. Since $v$ is indeed up to a
constant uniquely determined by any GS $n$, the functional derivative of
this $F$ exists at least for $n \in {\cal A}_{\text{DM},N}$ as a
derivative (`gradient', more precisely Frechet derivative) in the
hyperplane $\{n \in X \,|\, \langle 1,n \rangle = N\}$.

Less clear is the situation in spin DFT.\cite{ep01} Now, also $F$ need
not have a derivative for GS densities.

\section{A Few Essentials on Legendre Transforms}

Let $X = X^{**}$ and $X^*$ be two mutually dual functional spaces, that
is, $X^*$ comprises all norm-continuous linear functions on $X$ and vice
versa. ($(\mathbb R^N)^* = \mathbb R^N$ is the space of all gradient
vectors to functions on $\mathbb R^N$; $(\bm L^3(\mathbb T^3))^* = \bm
L^{3/2}(\mathbb T^3)$ and vice versa.) The Legendre transform $f^*(n),\;
n \in X$ of a function $f(v),\; v \in X^*$ is defined as
\begin{equation}
  \label{eq:III01}
  f^*(n) = \sup_{v \in X^*} \{\langle n,v \rangle - f(v)\}.
\end{equation}
A second Legendre transformation yields
\begin{equation}
  \label{eq:III02}
  f^{**}(v) = \sup_{n \in X} \{\langle v,n \rangle - f^*(n)\}.
\end{equation}
All we need is
\begin{enumerate}
\item $f^*(n)$ is a convex function of $n$, no matter what $f(v)$ is; if
  $f(v)$ is convex, then $f^{**}(v) = f(v)$; in general $f^{**}(v) \leq
  f(v)$.
\item $f(v) + f^*(n) \geq \langle v,n \rangle$; if, for convex $f$,
  $f(v) + f^*(n) = \langle v,n \rangle$, then $v \in \partial f^*(n)$
  and $n \in \partial f(v)$.
\end{enumerate}
In the second statement $\partial f(v)$ is the subdifferential of the
convex function $f$ at point $v$: the set of all linear functions
$\langle n,v' \rangle$ so that $f(v') \geq f(v) + \langle n,(v'-v)
\rangle$ for all $v' \in X^*$. If this set consists of a single linear
function only, then this linear function is the (total) differential
$df(v)$, that is, $n$ is the derivative of $f$ at $v$.

To elucidate these properties one may consider convex functions of one
real variable, $f(N)$ and $f^*(\mu)$ (see e.g.\ Fig.~11 of
Ref.~\onlinecite{esc03}). Put a supporting tangent to the graph of $f$
at point $N$ (a line having the common point $(N,f(N))$ with the graph
of the function and being nowhere above). The tangent has a slope $\mu$.
The sign carrying distance from the intersection point of this line with
the $f$-axis to the coordinate origin is $f^*(\mu)$. If $f$ has a
derivative at $N$, then its value is $\mu$. It is easily seen that, if
the derivative of $f$ jumps at $N$, then there is a (closed) interval
$[\mu_1,\mu_2]$ from the left derivative $\mu_1$ to the right derivative
$\mu_2$ ($\mu_1$ may be $-\infty$ or $\mu_2$ may be $+\infty$), and
$f^*(\mu)$ is \textit{linear} on this interval, the interval being the
subdifferential $\partial f(N)$. Inversely, if the convex function $f$
is not strictly convex, but has a linear dependence on some interval
with slope $\mu$, then the derivative of $f^*$ jumps at that $\mu$.

This simple geometric picture readily transfers to the general case:
take a tangent hyperplane $f(n_0) + \langle v,(n - n_0) \rangle$
supporting the graph of $f(n)$ at some $n_0$. The distance from its
intersection point with the $f$-axis to the origin is $f^*(v)$. If the
derivative of $f$ jumps at some $n$ (and hence does not exist there),
then there is a convex domain in $v$-space on which $f^*(v)$ is linearly
depending on $v$, and vice versa.

If the GS wave function is independent of some potential change $\delta
v$ called a `phantom' potential perturbation in Ref.~\onlinecite{koh04},
then the GS density $n$ does also not change and the GS energy has a
linear dependence $\text{const.} + \langle \delta v,n \rangle$.
Consequently, the functional derivative of $F[n]$ defined by
(\ref{eq:II16}) does not exist at that $n$. This is precisely the role
of `phantom' potential perturbations in DFT.

\section{Unique Mappings for $T > 0$}

We now move to temperature $T > 0$ and to grand canonical states. We
also generalize to spin DFT and allow for external magnetic fields
coupling to the particle spin but not to its charge (diamagnetic
couplings as usually in spin DFT are neglected). Consider a system of
identical particles in an external field $v_{ss'}(\bm r)$. Let the
system be confined in a large box, or, placed in a large three-torus
equivalent to periodic boundary conditions (regular $\bm k$-grid). Let
the Hamiltonian be that of (\ref{eq:II01}-\ref{eq:II03}), but
(\ref{eq:II04}) generalized to
\begin{equation}
  \label{eq:IV04}
  \hat V = \sum_{ss'}\int 
  \hat\psi^\dagger(\bm r,s)v_{ss'}(\bm r)\hat\psi(\bm r,s')\, d^3r,
\end{equation}
The particle number operator is $\hat N = \int
\hat\psi^\dagger(x)\hat\psi(x)\, dx$ so that $\hat H - \mu\hat N$
depends on the combination $v-\mu = v_{ss'}(\bm r) - \mu\delta_{ss'}$
only.

Fix the temperature $\beta = 1/kT$, the chemical potential $\mu$ and the
external potential $v$. Then, the grand canonical state is
\begin{equation}
  \label{eq:06}
  \rho_\beta[v-\mu] = \frac{e^{-\beta(\hat H - \mu\hat N)}}{\text{tr}\, 
  e^{-\beta(\hat H - \mu\hat N)}}.
\end{equation}
If $\rho > 0,\; \text{tr}\,\rho = 1,$ is \emph{any} state (density
matrix), then $\text{tr}\, \rho\, (\hat V - \mu\hat N) = 
\int (v-\mu)n[\rho]\, dx$ with the particle (spin) density
\begin{equation}
  \label{eq:07}
  n[\rho] = n_{ss'}(\bm r) = \text{tr}\, 
  \rho\,\hat\psi(\bm r,s)\hat\psi^\dagger(\bm r,s').
\end{equation}
In the following $\text{tr}$ will always mean the trace in the
Fock space of the $\hat\psi$. Also, the natural abbreviation
\begin{equation}
  \label{eq:08}
  \sum_{ss'}\int (v_{ss'}(\bm r) - \mu\delta_{ss'})n_{s's}(\bm r) d^3r =
  \langle (v-\mu),n \rangle
\end{equation}
will be used.

Now, fix the particle interaction $w$ and, following Mermin\cite{mer65}
(we try carefully to trace functional dependences and in doing so
slightly deviate from Mermin's notation), consider for various external
potentials $v$ the functionals
\begin{equation}
  \label{eq:09}
  \Omega_v[\rho] = \text{tr}\, \rho\,\biggl(\hat H - \mu\hat N + 
  \frac{1}{\beta}\ln\rho\biggr).
\end{equation}
As easily seen by direct substitution of (\ref{eq:06}), the grand
canonical potential $\Omega_\beta[v-\mu]$ is obtained as
\begin{equation}
  \label{eq:10}
  \Omega_\beta[v-\mu] = -\,\frac{1}{\beta}\ln\text{tr}\, 
  e^{-\beta(\hat H - \mu\hat N)} = \Omega_v[\rho_\beta[v-\mu]].
\end{equation}
Moreover, as shown in Ref. \onlinecite{mer65}, for any $\rho > 0,\;
\text{tr}\, \rho = 1$, it holds that
\begin{equation}
  \label{eq:11}
  \Omega_v[\rho] > \Omega_v[\rho_\beta[v - \mu]] = 
  \Omega_\beta[v - \mu] \text{ for } \rho\neq\rho_\beta[v-\mu].
\end{equation}
In Mermin's approach, this inequality replaces the corresponding ground
state energy property. It immediately implies that $\Omega_\beta[v-\mu]
= \min_\rho \Omega_v[\rho]$ is concave in $v$ by the simple reasoning
(we write in short $v_i$ for $v_i-\mu_i$)
\begin{equation}
  \label{eq:12}
  \begin{split}
    \lefteqn{\Omega_\beta[\alpha v_1 + (1-\alpha)v_2] \;=}\\ 
    &= \min_\rho \text{tr}\, \rho\,\biggl(\alpha\hat H_1 + 
    (1-\alpha)\hat H_2 + \frac{1}{\beta} \ln \rho\biggr) \;\geq\\
    &\geq \alpha\min_{\rho_1} \text{tr}\, \rho_1
    \biggl(\hat H_1 + \frac{1}{\beta} \ln \rho_1\biggr) \;+\\
    &\qquad (1-\alpha)\min_{\rho_2} \text{tr}\, \rho_2
    \biggl(\hat H_2 + \frac{1}{\beta} \ln \rho_2\biggr) \;=\\
    &= \alpha\Omega_\beta[v_1] + (1-\alpha)\Omega_\beta[v_2],
    \qquad 0 \leq\alpha\leq 1,
  \end{split}
\end{equation}
because a joint minimum of a sum cannot be below the sum of the
independent minima of the items.

As another advantage over the standard zero temperature theory, it
follows immediately from (\ref{eq:06}) and (\ref{eq:07}) that the
mappings $(v - \mu) \mapsto \rho_\beta \mapsto n$ are unique. There is
no problem with degenerate states since degenerate states automatically
get equal statistical weight in $\rho_\beta$ of (\ref{eq:06}). However,
as usual spontaneous symmetry breaking is not covered by this
statistical approach; it has to be treated by an infinitesimal symmetry
breaking external potential $v$ in the spirit of Bogolubov's quasi-means
in Statistical Physics. Nevertheless, by virtue of (\ref{eq:11}) which
also holds in the spin case, in the standard way Mermin proved the
analogue of the Hohenberg-Kohn lemma: $n \mapsto (v - \mu)$ is unique
for any $n$ coming from a grand canonical ensemble at temperature
$1/k\beta$. In summary, there are the unique mappings
\begin{equation}
  \label{eq:13}
  \begin{CD}
  (v - \mu) @<<< n\\
  @VVV           @AAA\\
  \text{\makebox[2em][l]{\hspace{-0.5em}grand canonical $\rho_\beta$}}\\
  \end{CD}
\end{equation}
On the functional domain (which may depend on $\hat W$)
\begin{equation}
  \label{eq:14}
  D_\beta = \big\{n \text{ coming from some } 
  \rho_\beta,\; \beta \text{ fixed}\big\}
\end{equation}
one can write $(v - \mu)_\beta[n]$ and also $\hat H_\beta[n]$ and
$\rho_\beta[n]$, as well as $n_\beta[v-\mu] = n[\rho_\beta[v-\mu]]$ on
the domain of admissible potentials $v$. (Denoting the distinct
functions $\rho_\beta[n]$ and $\rho_\beta[v]$ by the same symbol
$\rho_\beta$ will cause no confusion.)

Moreover, from the unique dependence of $v-\mu$ on $\rho_\beta$ it
follows now also that $\Omega_v[\rho]$ for different $v_i - \mu_i$ is
minimized by different $\rho_i$, and hence the inequality in
(\ref{eq:12}) is sharpened into a strict inequality: $\Omega_\beta[v]$
is strictly concave. If equality would hold in (\ref{eq:12}), this would
imply that the minimizing $\rho$ is also a minimizing $\rho_1$ for $v_1$
and a minimizing $\rho_2$ for $v_2$. This is the principal difference
from the $T = 0$ theory where $E[v,N]$ is not always strictly concave in
$v_{ss'}$ and is never strictly convex in $N$.

\section{The Density Functional}

As was already said, for electron systems it is well justified to allow
for all potentials
\begin{equation}
  \label{eq:16}
  v - \mu \in \bm L^{3/2}(\mathbb T^3) = X^*
\end{equation}
for which the integral $\int_{\mathbb T^3} |v - \mu|^{3/2}\, d^3r$ over
the three-torus (of finite volume) is finite. Recall that the
Hamiltonian $\hat H^0 = \hat T + \hat V$ of interaction-free fermions is
bounded below for any such potential, that this also holds true for
Hamiltonians (\ref{eq:II01}-\ref{eq:II03}, \ref{eq:IV04}), if $w(\bm
r,\bm r') \geq 0$, and that, since the space $\mathbb T^3$ has finite
volume, all considered Hamiltonians have discrete spectra with at most
finite degrees of level degeneracy. Then, $\Omega_\beta[v - \mu]$ of
(\ref{eq:10}) is well defined on $X^*$ and smooth in the norm topology.

In view of the concavity of $\Omega_\beta[v]$, introduce the Legendre
transform\cite{lie83,esc03} of $-\Omega_\beta[v]$ as $\tilde
F_\beta[-n]$:
\begin{equation}
  \label{eq:17}
  F_\beta[n] = \tilde F_\beta[-n] = 
  \sup_v \Bigl\{ -\langle n,v \rangle + \Omega_\beta[v]\Bigr\}
\end{equation}
which as a Legendre transform is a convex functional of $-n$ (or
likewise of $n$), the dual variable to $v$: $n \in X^{**} = X = \bm
L^3(\mathbb T^3)$.

Since the functional space $X$ is reflexive, $\bm L^3(\mathbb T^3) =
(\bm L^3(\mathbb T^3))^{**}$, the Legendre back transformation from
(\ref{eq:17}), $-\Omega_\beta[v] = \sup_n \{-\langle v,n \rangle -
\tilde F_\beta[-n]\}$ or,
\begin{equation}
  \label{eq:18}
  \Omega_\beta[v] = 
  \inf_n \Bigl\{F_\beta[n] + \langle v,n \rangle\Bigr\}
\end{equation}
represents the generalized Hohenberg-Kohn theorem (where equality holds
since $\Omega_\beta[v]$ is concave in $v$). The chemical potential $\mu$
is further on put to zero which simply means that single particle
energies and potentials are measured from the chemical potential.

For any density $n \in D_\beta$ from (\ref{eq:14}), in analogy to the
original Hohenberg-Kohn functional one may define
\begin{equation}
  \label{eq:19}
  \Phi_\beta[n] = \Omega_v[\rho_\beta[n]] - \langle v,n \rangle, \quad
  n \in D_\beta
\end{equation}
and, from (\ref{eq:11}), have
\begin{equation}
  \label{eq:20}
  \Omega_\beta[v] = \min_{n \in D_\beta} 
  \Bigl\{\Phi_\beta[n] + \langle v,n \rangle\Bigr\}
\end{equation}
since in view of (\ref{eq:13}) any $n \neq n[v]$ refers to
$\rho_\beta[n] \neq \rho_\beta[n[v]]$. From (\ref{eq:18}) and
(\ref{eq:20}) one infers that $F_\beta[n] = \Phi_\beta[n]$ for all $n
\in D_\beta$, and that the infimum of (\ref{eq:18}) is always a minimum
with minimizing density $n_\beta[v]$: $\Phi_\beta[n_\beta[v]] + \langle
v,n_\beta[v] \rangle = \Omega_\beta[v] \leq F_\beta[n_\beta[v]] +
\langle v,n_\beta[v] \rangle$, hence $\Phi_\beta[n] \leq F_\beta[n]$ for
$n \in D_\beta$, and by interchanging the role of (\ref{eq:18}) and
(\ref{eq:20}) in the argument the opposite inequality is obtained.

Moreover, for $-\Omega_\beta[v]$ and $\tilde F_\beta[-n]$ like in
general for any pair of mutual Legendre transforms it holds that $\tilde
F_\beta[-n] - \Omega_\beta[v] = -\langle n,v \rangle$
implies $v \in \partial\tilde F_\beta[-n]$ and $-n \in
\partial(-\Omega_\beta[v])$, where $\partial\tilde F_\beta[-n]$ means
the subdifferential on $\tilde F_\beta$ at point $-n$ and
$\partial(-\Omega_\beta[v])$ means the subdifferential on
$-\Omega_\beta$ at point $v$. Since $d\Omega_\beta[v] =
\ln\text{tr}\,(e^{-\beta\hat H}d\hat V)$ as easily seen from the
definition of $\Omega_\beta[v]$, its first (Frechet) derivative (for
finite volume $|\mathbb T^3|$) always exists so that its subdifferential
contains only this one `gradient'. Now, the reasoning after
(\ref{eq:20}) yields $F_\beta[n_\beta[v]] - \Omega_\beta[v] = -\langle
n_\beta[v],v \rangle$ and hence
\begin{equation}
  \label{eq:22}
  \frac{\delta\Omega_\beta}{\delta v} = n_\beta[v].
\end{equation}
As $n_\beta[v]$ is a one-one mapping $X^* \leftrightarrow D_\beta$,
for $n \in D_\beta$ one has inversely $v_\beta[n]$ and
$F_\beta[n] - \Omega_\beta[v_\beta[n]] = -\langle n,v_\beta[n] \rangle$
implying
\begin{equation}
  \label{eq:23}
  \frac{\delta F_\beta}{\delta n} = -v_\beta[n], \qquad
  n \in D_\beta.
\end{equation}
Note that while (\ref{eq:23}) holds for $n \in D_\beta \subset X$, the
derivative $\delta/\delta n$ on the left hand side is taken in $X$, that
is, for \textit{any} $\delta n \in X$ with $\|\delta n\|$ small enough.

From the strict concavity and continuous differentiability of
$\Omega_\beta[v]$ the differentiability of $F_\beta[n]$ at every point
$n \in D_\beta$ follows, that is, at every density $n$ thermodynamically
corresponding to some $v$ at temperature $(k\beta)^{-1}$. Like in the $T
= 0$ theory\cite{lie83,esc03}, (\ref{eq:17}) yields that $F_\beta[n]$
jumps to $+\infty$ if $n < 0$ for an $x$-domain of non-zero measure:
Take $v = c > 0$ for some domain where $n < 0$ and $v = 0 (= \mu)$
everywhere else.  This $v$ is admissible for arbitrary large $c$ and
$\hat H[v]$ is bounded below for such a $v$. Hence, $\Omega_\beta[v]$ is
also bounded below and, as easily seen, the supremum (\ref{eq:17}) is
obtained for $c \rightarrow \infty$ to be $+\infty$. Assume now that
$n_\beta(x_0) = 0$ for some $x_0$. Since any $n_\beta \in D_\beta$ is
continuous in $x$ (any solution of the many-particle Schr\"odinger
equation is continuous), there is always $\delta n \in X,\; \delta
n(x_0) > 0,$ so that $n_\beta(x) - \epsilon\delta n(x)$ would be
negative in a neighborhood of $x_0$ of non-zero measure for arbitrarily
small $|\epsilon|$ and the functional derivative (\ref{eq:23}) would not
exist for that $n_\beta$. Thus, the result (\ref{eq:23}) also implies
$n_\beta(x) > 0$ everywhere for $T > 0$.  (See also next section.)

\section{Interaction-free Particles and Beyond}

As is well known from Statistical Physics\cite{lan80}, in a
non-interacting particle system the particles in a single particle
quantum state $|\varphi_k\rangle$ may be treated as an independent
subsystem even of a quantum ensemble with exchange symmetry.  The
corresponding statistical fermionic state is
\begin{equation}
  \label{eq:24}
  \rho_k = |\rangle(1-p_k)\langle| + 
  |\varphi_k\rangle p_k \langle\varphi_k|
\end{equation}
with occupation $p_k$ of the orbital $\varphi_k$ and $|\rangle$ as the
vacuum state. Accordingly we define\cite{ww92}
\begin{equation}
  \label{eq:25}
  \begin{split}
  {\cal T}_\beta[p_k,\varphi_k] &= \text{tr}\; \rho_k 
  \biggl( \hat T + \frac{1}{\beta} \ln\rho_k \biggr)\; = \\
  &= -p_k (\varphi^*_k\,|\hat t|\,\varphi_k) +\\ 
  &+ \frac{1}{\beta}\,\bigl(p_k \ln p_k + (1-p_k) \ln(1-p_k)\bigr)
  \end{split}
\end{equation}
and the density functional
\begin{equation}
  \label{eq:26}
  T_\beta[n] = \min_{\substack{\{p_k,\varphi_k\}\\0 \leq p_k \leq 1\\
  (\varphi_k|\varphi_{k'}) = \delta_{kk'}}}
  \biggl \{\sum_k {\cal T}_\beta[p_k,\varphi_k]\biggm|
  \sum_k p_k|\varphi_k|^2 = n \biggr\}
\end{equation}
where the minimum taken over all orthonormal orbitals and orbital
occupations which yield a given $n$ exists like in the GS case.

Now, the grand canonical potential is
\begin{equation}
  \label{eq:27}
  \begin{split}
    \lefteqn{\Omega^0_\beta[v] 
    = \min_n \biggl\{T_\beta[n] + \langle v,n \rangle\biggr\}\; =} \\
    &= \min_{\substack{\{p_k,\varphi_k\}\\0 \leq p_k \leq 1\\
       (\varphi_k|\varphi_{k'}) = \delta_{kk'}}}
       \biggl\{\sum_k {\cal T}_\beta[p_k,\varphi_k] + 
       \langle v,n[p_k,\varphi_k] \rangle\biggr\}
  \end{split}
\end{equation}
where again the $\varphi_k$ must be orthonormal.

Variation of the $\varphi^*_k$ under the last constraint yields for the
minimizing orbitals $\varphi^0_k$
\begin{equation}
  \label{eq:28}
  \bigl(\hat t + v \bigr)\varphi^0_k = 
  \varphi^0_k\varepsilon^0_k\,,
\end{equation}
and variation of the $p_k$ yields
\begin{equation}
  \label{eq:29}
  p^0_k(\beta) = f_\beta(\varepsilon^0_k) = 
  \frac{1}{e^{\beta\varepsilon^0_k} + 1}\,,
\end{equation}
which is the correct result in this physically trivial case. For any $v
\in X^*$ the minimum of (\ref{eq:27}) does indeed exist, and the
minimizing density is
\begin{equation}
  \label{eq:30}
  n^0_\beta[v] = n^0_{\beta ss'}(\bm r) = 
  \sum_k f_\beta(\varepsilon^0_k)
  \varphi^0_k(\bm r,s)\varphi^{0*}_k(\bm r,s')
\end{equation}
so that $N = \sum_k f_\beta(\varepsilon^0_k)$
relates the average particle number $N$ to the chemical potential $\mu$.
Only the occupation numbers depend on temperature $(k\beta)^{-1}$ and on
the value of the chemical potential $\mu$ from which $v$ and
the $\varepsilon^0_k$ are measured.

Here, $n^0_\beta > 0$ everywhere is intuitively clear because
$f_\beta(\varepsilon^0_k) > 0$ for all $k$.

For densities minimizing (\ref{eq:27}) it obviously holds that
\begin{equation}
  \label{eq:32}
  T_\beta[n] = \Phi^0_\beta[n] = F^0_\beta[n], \qquad n \in D^0_\beta.
\end{equation}
($T_\beta[n]$ replaces the density matrix functional $T_{\text{DM}}[n]$
of the ground state theory, Eq.~(\ref{eq:II09}).)

Accounting for the Coulomb interaction of the electrons in mean-field
approximation simply means to replace $v$ in the above Schr\"odinger
equation by $v + v^\text{H}$ where
\begin{equation}
  \label{eq:33}
  v^\text{H}(\bm r) = \int 
  \frac{n_{\beta s's'}(\bm r')}{|\bm r - \bm r'|}\, dx'
\end{equation}
contains self-interaction. Since\cite{lie83} $\nabla\varphi_k \in \bm
L^2(\mathbb T^3)$ implies $|\varphi_k|^2 \in \bm L^3(\mathbb T^3)$,
taken as a KS ansatz $n = \sum_k p_k|\varphi_k|^2 \geq 0$ is
sufficiently general for the density of an interacting system too.
Densities of this type apparently form a convex domain $D$ of the
functional space $X = \bm L^3(\mathbb T^3)$ on which $T_\beta[n]$ is
also defined by (\ref{eq:26}). (${\cal J}_N \subset D$ for every real
$N$, $0 \leq N < \infty$.)

As $D_\beta \subset D$ also for $\hat W \neq 0$, by
\begin{equation}
  \label{eq:34}
    F_\beta[n] = T_\beta[n] + \frac{1}{2}
    \int\frac{n_{ss}(\bm r)n_{s's'}(\bm r')}{|\bm r-\bm r'|}
    \, dxdx' + F^{\text{XC}}_\beta[n]
\end{equation}
for $n \in D$ an exchange and correlation density functional
$F^{\text{XC}}_\beta[n]$ is defined (since the other density functionals
of this relation were previously defined or are explicitly given on
$D$).  Inserting here $n = \sum_k p_k|\varphi_k|^2$ transforms
(\ref{eq:18}) into a minimum search by varying $\varphi^*_k$ and $p_k$
as above in the GS theory. The derivatives with respect to $\varphi^*_k$
and $p_k$ of $F_\beta$ exist on the basis of (\ref{eq:23}) for
$\{p_k,\varphi_k\}$ yielding $n \in D_\beta$, and those of the second
term on the right hand side of (\ref{eq:34}) are explicitly known.
Hence, the situation with $T_\beta$ and $F^{\text{XC}}_\beta$ is like in
the GS theory. We cannot expect $D_\beta \subset D^0_\beta$.

Formally, like in (\ref{eq:II14}) one may again assume
\begin{equation}
  \label{eq:35}
  \begin{split}
  \lefteqn{T_\beta[n] + F^{\text{XC}}_\beta[n] =}\\
  &\min_{\substack{\{p_k,\varphi_k\}\\0 \leq p_k \leq 1\\
        (\varphi_k|\varphi_{k'}) = \delta_{kk'}}}
  \Bigl\{ \sum_k {\cal T}_\beta[p_k,\varphi_k] + 
  {\cal F}^{\text{XC}}_\beta[\{p_k,\varphi_k\}] \Bigm|\\[-3ex]
  &\qquad\qquad\qquad\qquad\qquad\qquad
  \Bigm| \sum_k p_k |\varphi_k|^2 = n \Bigr\}
  \end{split}
\end{equation}
which yields the KS equation
\begin{equation}
  \label{eq:36}
  \bigl(\hat t + \hat v^{\text{eff}} - 
  \varepsilon_k\bigr) \varphi_k = 0,
\end{equation}\vspace{-1ex}
with
\begin{equation}
  \label{eq:37}
  \hat v^{\text{eff}}\varphi_k = 
  \bigl(v + v^\text{H} + \hat v^{\text{XC}}\bigr)\,\varphi_k, \quad
  \hat v^{\text{XC}}\varphi_k = \frac{1}{p_k}\,
  \frac{\delta{\cal F}^{\text{XC}}_\beta}{\delta\varphi^*_k}\,,
\end{equation}
and $p_k(\beta) = f_\beta$ from (\ref{eq:29}) with $\varepsilon^0_k$
replaced by $\varepsilon_k$. Compare also the previous discussion of the
property (\ref{eq:II15}) of $\hat v^{\text{XC}}$. Note that we did again
not prove the existence of $\delta F^{\text{XC}}_\beta/\delta n$:
$v^{\text{eff}}$, if it exists at all, need not exist as an orbital
independent local potential, it might be non-local and orbital
dependent. In this respect the situation is the same as for the ground
state theory.

Given the external potential $v$, the solutions of this KS equation
determine, via the analogues of (\ref{eq:29}, \ref{eq:30}) without
superscripts, the density $n_\beta[v] \in D_\beta$ minimizing the right
hand side of (\ref{eq:18}) and hence providing the grand canonical
potential
\begin{equation}
  \label{eq:38}
  \Omega_\beta[v - \mu] = F_\beta[n_\beta[v - \mu]] + 
  \int (v - \mu)n_\beta[v - \mu]\, dx.
\end{equation}
where we explicitly reinserted the chemical potential $\mu$. The latter
is related to the particle number $N$ by
\begin{equation}
  \label{eq:39}
  -\,\frac{\partial\Omega}{\partial\mu} = N = 
  \sum_k f_\beta(\varepsilon_k - \mu)
\end{equation}
which is also confirmed by inserting (\ref{eq:23}) and the KS expression
for $n$ into (\ref{eq:38}).

The whole theory, of course, as in the ground state variant again
depends on the knowledge of the density functional
$F^\text{XC}_\beta[n]$ and of ${\cal F}^{\text{XC}}_\beta$ in the KS
theory, both of which are hardly ever accessible (if the latter exists
at all) and hence have to be modeled changing the exact theory into a
model theory within a (nearly) rigorous frame. An early \textit{ad-hoc}
application is\cite{gha78}. Since (\ref{eq:39}) rests on (\ref{eq:23}),
it can be used as a check for the quality of a model
$F^\text{XC}_\beta[n]$, for instance down to which temperature it can
reasonably be used for a specific answer.

\acknowledgements{I am grateful to K. Koepernik for helpful
  discussions.}

\end{document}